\documentstyle[12pt]{article}
\newskip\humongous \humongous=0pt plus 1000pt minus 1000pt
  \newif\ifdtup

\def\ltap{\raisebox{-.4ex}{\rlap{$\sim$}} \raisebox{.4ex}{$<$}}

\def\Nto1{\raisebox{-1ex}{\rlap{\tiny $\;\;N \to 1$}} \raisebox{0ex}
{$\;\;\;\,\to\;\;\;\,$}}

\def\frac#1#2{ {{#1} \over {#2} }}

\def\as{\alpha_S}

\def\beq{\begin{equation}}
\def\eeq{\end{equation}}

\def\np#1#2#3{Nucl.\ Phys.\ B#1 (19#3) #2}
\def\pl#1#2#3{Phys.\ Lett.\ #1B (19#3) #2}

\def\prep#1#2#3{Phys.\ Rep.\ #1 (19#3) #2}

\def\rmp#1#2#3{Rev.\ Mod.\ Phys.\ #1 (19#3) #2}

\def\zp#1#2#3{Zeit.\ Phys.\ C#1 (19#3) #2}

\def\figcap{
        {\bf\large Figure Captions\markboth
        {FIGURECAPTIONS}{FIGURECAPTIONS}}\list
        {Fig. \arabic{enumi}:\hfill}{\vspace {4 mm}\settowidth
        \labelwidth{Figure 999:}
        \leftmargin\labelwidth
        \advance\leftmargin\labelsep\usecounter{enumi}}}
 \relax

\begin{document}
\par \vskip 10mm
\begin{center}
{
\Large \bf
Small $x$ processes: Heavy Quark\\
production at high energy}
\footnote{Research supported in part by MURST, Italy.
Talk given at the XXVIIth Rencontre de Moriond, March 1993,
Les Arcs, France
}
\end{center}
        \par \vskip 3mm \noindent
\begin{center}
        \par \vskip 3mm \noindent
        { GIUSEPPE MARCHESINI}\\
        \par \vskip 2mm \noindent
        Dipartimento di Fisica, Universit\`{a} di Parma,\\
        INFN, Gruppo Collegato di Parma, ITALY.\\
\end{center}
\par \vskip .7 true cm
\begin{center} {\large \bf Abstract} \end{center}
\begin{quote}
A brief summary is given of some recent perturbative QCD
results on the evaluation of cross sections and on the the
structure of final states in processes with incoming hadrons
at small $x$.
A new Monte Carlo simulation which includes these theoretical
features is described together with its applications to the study
of heavy quark production in $ep$ collisions at very high energy.

\end{quote}
\vspace*{\fill}
\renewcommand{\thefootnote}{\fnsymbol{footnote}}
Heavy flavour production is a process with a hard scale  given by
the quark mass $M$. At very high energies this scale is much smaller
than the collision c.m.\ energy $\sqrt s$.
This implies that together with the collinear logarithms (powers of
$\ln M^2/\Lambda^2$) we must resum also powers of $\ln x$ with
$x \simeq M^2/s \ll 1$.
Much progress has been made over the past few years in the theoretical
understanding of small-$x$ processes.
In particular the key acievements in the region $x \to 0$ are:
(i) a better understanding of the ``Lipatov'' anomalous dimension
\cite{Lip,CCFM};
(ii) the resummation \cite{CCH} to all loop order the
leading contributions in the coefficient function;
(iii) the extension to this region of the coherent branching process;
(iv) the possibility to resum all these new contributions by Monte
Carlo methods \cite{Smallx,LMRW,HFEP}.
In this talks I will briefly summarize these theoretical results
and describe a recent application of the Monte Carlo simulation
to heavy flavour leptoproduction at Hera and higher energies.

\vskip 0.3 true cm \noindent
1) {\it Structure function.}
\vskip 0.2 true cm \noindent
The structure  function is given in term of the space-like
anomalous dimension  $\gamma^S_N(\as)$
(the limit $x \to 0$ corresponds to $N \to 1$, where $N$ is the
energy moment index).
The leading contributions in $\gamma^S_N(\as)$ are given by an
expansion in powers of $\as/(N-1)$ known since long time \cite{Lip} and
recently studied in the framework of hard processes \cite{CCFM}.
The first terms of the ``Lipatov'' anomalous dimension are
\begin{equation}
\label{gammae}
\gamma^S_N (\as) =
\frac {\bar \as} {N-1}
+ 2 \zeta_3 \left(\frac {\bar \as} {N-1}\right)^4
+ 2 \zeta_5 \left(\frac {\bar \as} {N-1}\right)^6
+12 \zeta_3^2 (\frac {\bar \as} {N-1})^7
+ \cdots
\end{equation}
where $\bar \as = C_A \as/\pi$ and
$\zeta_i$ is the Riemann zeta function.
There ere are no leading terms of order $\as^2$, $\as^3$, and $\as^5$.
Although each term is singular only at $N=1$, this expansion
develops a square root singularity at $N = 1+ (4 \ln 2 ) \bar \as$.
The presence of this singularity at $N>1$ implies
that the behaviour of the structure function for $x \to 0$ is more
singular than that given by any finite number of loops.
For fixed $\as$ and small $x$ the behaviour of the one loop
structure function is
\begin{equation}
\label{1loop}
xF^{(1)}(x,Q) \sim \exp \sqrt{a\ln(1/x)}\;,
\;\;\;\;\;\;\;\;\;\;\; a = 4\bar\as\ln(Q^2/Q^2_s)\; .
\end{equation}
By summing the all loop result in Eq.~(\ref{gammae}) one finds
instead the following behaviour
\begin{equation}
\label{Lipf}
xF^{(all)}(x,Q) \sim x^{-p}\;,\;\;\;\;\;\; p=(4 \ln 2)\bar\as\; ,
\end{equation}
which is much more singular for $x\to 0$.

In spite of this quite different behaviour, it turns out that at
this highly inclusive level the all-loop and the conventional
one-loop formula give similar results \cite{Smallx}.
This is partially due to the fact that the first correction to the
one-loop expression of the anomalous dimension is to order $\as^4$.
Thus the steeper behaviour of the structure function is seen only
for very low $x$.
Moreover it has been pointed \cite{LMRW} that the steeper behaviour
for $\as$ running is even more asymptotic than for fixed $\as$.
This is due to the presence of the cutoff in the exchanged transverse
momenta $k_{ti}>Q_0$. Although this condition is asymptotically negligible,
it has some effect in reducing the evolution of the
branching in the first steps. As a result the distribution at small
$x$ is somewhat reduced. For a detailed discussion see \cite{LMRW}
and the contribution by Ryskin to this conference.

\vskip 0.3 true cm \noindent
2) {\it Cross section.}
\vskip 0.2 true cm \noindent
To compute the heavy flavour production cross section
one convolutes the structure function with the coefficient function.
Recently a way has been found to sum all leading terms
$\as^n/(N-1)^n$ in the coefficient function as well.
The key development \cite{CCH} in this study has been
a generalized factorization theorem in which one takes into
account the off-shellness of the hard scattered parton.
One introduces the hard elementary off-shell cross section
$\hat \sigma(k_t,Q) $ for a photon and a gluon  to produce a
heavy quark-antiquark pair, where $q^2=-Q^2$ and $k^2=-k_t^2$ are
the photon and gluon squared masses respectively,
and the generalized proton structure function
${\cal F}(x,\mu,k_{t})$
giving the probability (per unit of $\ln x$) of finding
a gluon at longitudinal momentum fraction $x$ and transverse
momentum $k_t$ in a hard process at the scale $\mu$.
Integrating this distribution over $k_t<\mu$ one obtains
the gluon structure function $F(x,\mu^2)$.
By studying  $\hat \sigma(k_t,Q)$ at Hera energy and $M=5$GeV
one has \cite{HFEP} that when $W^2\gg M^2\gg Q^2 \gg \Lambda^2$,
$\; W$ being the hadronic c.m.\ energy, the natural cutoff is
around $4M^2$, while for deep inelastic production with
$W^2\gg Q^2\gg M^2$ the cutoff becomes $Q^2$.
Therefore the hard scale is typically assumed to be $\mu^2=4M^2+Q^2$.
However at small $Q^2$ or at $W \simeq M$, the dynamical suppression
in $k_t$ is at a significantly smaller scale.

In the conventional calculation, the heavy flavour cross section is
obtained by convoluting the on-shell elementary cross section
$\hat \sigma(0,0)$ and the gluon structure function $F(x,\mu^2)$.
This procedure has two main effects:
(i) for $k_t^2<\mu^2$ the elementary cross section is overestimated by
its on-shell value $\hat \sigma(0,0)$;
(ii) the `tail' of the cross section at $k_t^2>\mu^2$ is ignored.

Asymptotically, the second effect dominates and the
cross section is expected \cite{CCH} to be larger than
the conventional on-shell Born approximation.
At subasymptotic energies the first effect is important
and one overestimates the cross section.
The Monte Carlo simulation based on the coherent branching algorithm
\cite{HFEP}
predicts that the b-quark leptoproduction cross section at Hera
energy is lower than the one obtained by a conventional one-loop
calculation \cite{EK}.

\vskip 0.3 true cm \noindent
3) {\it Coherent branching for $x \to 0 $ and $x \to 1$.}
\vskip 0.2 true cm \noindent
It has been recently shown \cite{CCFM} that in the small $x$
region one can resum the leading contributions of gluon emission
by a branching algorithm which has the following two characteristics.

(i) {\it Phase space for the branching}.
Destructive interference among soft gluons  depletes the emission
phase space and one finds that both for large and small $x$
the emission takes place in the angular ordered region.
Denoting by $\theta_i$, $q_{ti}$ and $z_i$ the angle,
the transverse momentum with respect to the incoming hadron
and the exchanged energy fraction
of the $i$ emitted gluon we have, for small $z_i$,
\begin{equation}
\label{123}
\{ \theta_{i+1} > \theta_{i} \}
\;\sim\;\{ q_{t \, i+1} > z_{i} q_{t\, i} \} \,.
\end{equation}
For small $z_i$ this phase space is larger than the conventional
one corresponding to transverse momentum ordering and leading to
the one-loop anomalous dimension.

(ii) {\it Non-Sudakov form factor}.
In the region $x \to 0$ some of the gluons $q_i$ have $ z_i \to 0$.
The corresponding virtual corrections contain $\ln z_i$-singular
contributions, which factorize and exponentiate
to give the following non-Sudakov form factor
\begin{equation}
\label{ns}
\Delta_{ns}(z_i,q_{ti},k_{ti})\,=\,
= \exp \left[ -\frac{C_A }{\pi} \, \as(k_{ti}) \,
\ln \left( {\frac{1}{z_i}} \right)
\ln \left( \frac{k_{ti}^2 }{z_i q_{ti }^2} \right) \right] \, ,
\end{equation}
where $k_{ti}$ is exchanged transverse momentum resulting after
the emission of gluon $q_i$, i.e. is the total transverse momentum
of the system formed by all partons emitted within a cone of
aperture $\theta_i$.
This form factor becomes negligible for finite $x$ but for small
$x$ is important and has to be considered together with the usual
Sudakov form factor.

The non-Sudakov form factor has the effect
of screening the $1/z_i$ singularity of the gluon splitting function
\begin{equation}
\label{zto0}
\Delta_{ns}(z_i,q_{ti},k_{ti})\, / z_i  \to 0 \;\;\;\;\; z_i \to 0 \, .
\end{equation}
The $k_{ti}$-dependence
in $\Delta_{ns}$ makes the branching {\it non local}, i.e. dependent
on the development of part of the emission process. Because of this
non local $k_{ti}$-dependence, the new branching does not leads, for
$x \to 0$,
to the Altarelli-Parisi equation for the structure function. One
obtains instead the Lipatov equation for the structure function
which gives the ``Lipatov'' anomalous dimension.

Angular ordering is also the phase space constraints obtained from
coherence of soft radiation in the large $x$ region \cite{IR}.
Therefore it is possible to construct a unified coherent
branching algorithm valid both for $x \to 0 $ and $x \to 1$,
which for $x \to 0 $ takes into account the mentioned results
to all-loops, and for $x$ finite takes into account all the
leading contributions and the next-to-leading corrections important
in the large $x$ region. The important difference between the
time-like and the space-like branching is the presence in the
latter of the non-Sudakov form factor.

\vskip 0.3 true cm \noindent
4) {\it Unified equation for the structure function.}
\vskip 0.2 true cm \noindent
{}From the coherent branching algorithm valid both for small and
large $x$, one obtains \cite{MErice} a unified equation for the
generalized structure function ${\cal F}(x,Q,k_{t})$
giving the probability (per unit of $\ln x$) of finding
a gluon at longitudinal momentum fraction $x$ and transverse
momentum $k_t$ in a hard process at the scale $Q$. This hard scale
gives the maximum available angle for the branching.
Taking into account only the contributions which are singular for
$x \to 0$ and $1$, for the energy moment distribution one obtains
\beq
Q^2 \frac { \partial {\cal F}_N(Q,{\bf k}_{t}) }{\partial Q^2} =
\int dz \frac {\as C_A}{\pi}z^{N-1}
\left[ \left( \frac 1 {1-z} \right)_+
+ \frac 1 z \Delta_{ns}(z,\frac Q z ,k_{t})\, \right]
{\cal F}_N(\frac Q z ,{\bf k}_{t}-\frac{1-z}{z}{\bf Q} )\, .
\eeq
In the Kernel the emission of a gluon with energy fraction $z$ and
transverse momentum $q_t=(1-z)/z\, Q$ is factorized.
In the integrand we have the distribution before the emission of this
gluon with the total transverse momentum ${\bf k_t-q_t}$ and
at the scale $Q/z$ which is given by angular ordering.
The integration over the azimuthal direction of $q_t$ is understood.
The term $()_+$ is the usual $1/(1-z)$ singularity for the soft gluon
emission. Its regularization corresponds to the usual Sudakov form
factor.
For $N > 1$ we have $z={\cal O}(1)$ thus both the rescaling of $Q$
in ${\cal F}_N$ and the non-Sudakov form factor can be neglected and
the equation becomes the usual light cone expansion evolution equation.
The regular $z(1-z)-2$ term in the gluon splitting function is not
included in this approximation.
For  $N \to 1$ the rescaling of $Q$ and the screening of the $1/z$
singularity by the non-Sudakov form factor are important.
The equation has not any more the structure of an evolution equation
and becomes equivalent to the Lipatov equation and gives the anomalous
dimension in Eq.~(\ref{gammae}).

\vskip 0.3 true cm \noindent
5) {\it Monte Carlo simulation.}
\vskip 0.2 true cm \noindent
{}From the QCD coherent branching algorithm valid in all regions of $x$,
we have constructed \cite{Smallx,LMRW} a Monte Carlo simulation
program which has been applied \cite{HFEP} to the study of heavy
flavour leptoproduction at Hera and higher energies.
We have compared the results of this program, which for small $x$
reproduces the mentioned ``all-loop'' contributions, with the one
of the conventional ``one-loop'' branching, which for small $x$
reproduces only the first contribution of the anomalous dimension
in (\ref{gammae}). For large $x$ the two branchings are equivalent.
In the following I give a short summary of the main results of
Ref.~\cite{HFEP}.

In the study of heavy flavour leptoproduction, the most important
differences between the improved all-loop branching and the
conventional one-loop branching are seen in the final state
gluon distributions. These differences arise from the
additional phase space available for primary gluon emission in the
all-loop evolution: the region of disordered transverse momenta is
forbidden in the one-loop evolution, whilst in the new treatment it is
allowed, although suppressed at very small momentum fractions
by the non-Sudakov form factor. Therefore the number of emitted
gluons is enhanced, especially at small $x$ and large angles,
i.e.\ in the low-rapidity region.
At present, these differences are small compared with uncertainties due
to our lack of knowledge of the input gluon distribution. This underlines
the importance of determining the gluon structure function experimentally
down to the lowest possible values of $x$.

One of the most important feature of the new formulation is the
suppression of large energy and rapidity gaps, and large neighboring
pair masses, in the distributions of primary emitted gluons.
When the full colour structure of
final-state branching is taken into account, this will have the
effect of suppressing the production of high-mass colour-singlet
combinations of partons. This in turn will produce a more local
preconfinement of colour \cite{AV,LPHD} and permit a more direct
connection between the perturbative parton shower and the observed
hadron distributions.

The inclusive distributions generated by one-loop and all-loop
evolution are rather similar. This is mostly due to a cancellation
of leading higher-order corrections in inclusive observables, reflected
in the absence of leading singularities of the ``Lipatov''
anomalous dimension in second and third order and to the presence,
for running $\as$, of the cutoff in the exchanged transverse momenta.

The effects of the leading singularities of the coefficient function,
represented by the off-shell photon-gluon subprocess cross section,
are visible in the transverse momentum distribution of
the interacting gluon. However, the asymptotically dominant
effect of the high-$k_t$ tail is masked, even at very high energy,
by the lower value of the off-shell cross section at $k_t\ltap 2M$.
Although the dynamical simplifications prevent firm predictions
of the resulting cross sections, these results are confirmed
qualitatively by recent analytical studies of sub-asymptotic
effects \cite{CCH2}.

The program \cite{HFEP} used to generate these Monte Carlo results is
available from the authors. However, it should be emphasized
that the program is not an event generator: it produces weighted
events only and the weight distribution is broad. It also makes
numerous kinematical and dynamical simplifications.
Therefore it not suitable for detailed quantitative phenomenology,
but is intended rather as a theoretical tool to permit a first look
at exclusive small-$x$ phenomena.

\par \vskip 0.5 true cm

\end{document}